# Correlating surface stoichiometry and termination in SrTiO$_3$ films grown by hybrid molecular beam epitaxy

Suresh Thapa[1], Sydney R. Provence[1], Devin Jessup[2], Jason Lapano[3], Matthew Brahlek[3], Jerzy T. Sadowski[4], Petra Reinke[2], Wencan Jin[1], and Ryan B. Comes[1,a]

[1]Physics Department, Auburn University, Auburn, AL, 36849, USA

[2]Department of Materials Science and Engineering, University of Virginia, Charlottesville, VA 22904

[3]Materials Science and Technology Division, Oak Ridge National Laboratory, Oak Ridge, TN, 37830

[4]Center for Functional Nanomaterials, Brookhaven National Laboratory, Upton, NY, 11973

a) Electronic mail: ryan.comes@auburn.edu

**Abstract:** Hybrid oxide molecular beam epitaxy (hMBE), a thin-film deposition technique in which transition metal cations are delivered using a metal-organic precursor, has emerged as the state-of-the-art approach to the synthesis of electronic-grade complex oxide films with a stoichiometric growth window. However, numerous questions remain regarding the chemical mechanisms of the growth process and the surface properties of the resulting films. To examine these properties, thin film SrTiO$_3$ (STO) was prepared by hMBE using a titanium tetraisopropoxide (TTIP) precursor for Ti delivery and an elemental Sr source on annealed STO and Nb-doped STO substrates with varying TTIP:Sr flux ratios to examine the conditions for the reported stoichiometric growth window. The films were transferred *in vacuo* to an x-ray photoelectron spectroscopy system to study the surface elemental composition. Samples were examined using x-ray diffraction to compare our surface sensitive results with previously reported measurements of the bulk of the films in the literature. *Ex situ* studies by atomic force microscopy, scanning tunneling microscopy and low energy electron microscopy confirmed the presence of surface reconstructions and an Ehrlich-Schwoebel barrier consistent with an A-site SrO termination. We find that a surface exhibiting a mixture of SrO and $TiO_2$ termination, or a full SrO termination is necessary to obtain stoichiometric adsorption-controlled growth. These results indicate that surface Sr is necessary to maintain chemical equilibrium for stoichiometric growth during the hMBE process, which is important for the design of future interfacial systems using this technique.

## I. INTRODUCTION

Novel properties governed by the polar discontinuity at a heterojunction have been a central focus of oxide thin film research for over a decade. The electronic reconstruction at the interface due to the polar discontinuity between oxides layer can generate high mobility two-dimensional electron gas (2DEG)[1,2]. Atomic-scale control of interfacial terminations between polar and non-polar materials has led to a variety of emergent phenomena with potential applications for energy and electronic technologies. Beyond 2DEG systems, studies of polar/non-polar $LaFeO_3$/$SrTiO_3$ interfaces[3–5] and the subsequent demonstration of photocatalytic behavior[6] reflects the importance of understanding interfacial structures and defects in these materials with a great deal of precision for future electronic and energy devices. In order to engineer these materials, however, careful control over the synthesis process is required, which has led to the continued improvement of epitaxial growth techniques.

The epitaxial growth of atomic scale complex oxide thin films has been achieved by pulsed laser deposition (PLD) and molecular beam epitaxy (MBE) over the past 30 years. However, compared to the growth of traditional semiconductors, growth of complex oxides is challenging due to the lack of an adsorption-controlled growth window in MBE for most materials. Exceptions to this rule generally employ cations with high vapor pressures in either their metallic or oxide forms, such as Pb in $PbTiO_3$ or $RuO_x$ in $SrRuO_3$ [7]. In general, most cations do not exhibit such behavior, making it difficult to synthesize a material with perfect stoichiometry within current detection limits. Established oxide MBE employs effusion cells for each element, and recipes have been developed for the growth of $SrTiO_3$ (STO) and other materials that typically use alternating deposition of the *A*-site and *B*-site cations, allowing for stoichiometry control down to ~1% precision[8]. Achieving even more precise stoichiometric control, which is necessary for control of defect and dopant inventories, however, has proven challenging. Additional difficulties arise form the challenge to maintain stable evaporation rates for low vapor pressure and refractive elements in the requisite oxygen environment. These challenges can be overcome by using metal-organic precursors rather than elemental sources is a promising alternative to traditional oxide MBE. Hybrid MBE (hMBE) is a well-known technique for the implementation of mixed type sources combining elemental effusion cells for *A*-site cations and metal organic precursors for transition metal *B*-site cations in reactive oxygen sources.

Epitaxial growth for several refractory metal oxides, including vanadates, zirconates, and titanates, relies on hMBE[9–16]. Unexpectedly, hMBE was found to exhibit an adsorption-controlled growth window for stoichiometric STO growth when a titanium tetraisopropoxide (TTIP) precursor was used for titanium delivery in conjunction with evaporation of Sr from a conventional effusion cell[14]. The self-regulated growth window provides an avenue to control the stoichiometry of both the cations and anions which compensates unavoidable drift in effusion cell fluxes over the course of the growth. The presence of a growth window for STO via hMBE is now well-established and has led to the highest mobility $n$-STO films on record[13,17]. STO is one of the most widely studied perovskite oxide structures in terms of bulk and surface properties due to its unique properties along with the simple cubic structure. A stoichiometric adsorption-controlled growth window has also been reported for several other materials, including $SrVO_3$[15] and $BaSnO_3$[18], yielding some of the best electronic properties reported to date in these materials. However, evaluating the presence of a growth window for STO and other complex oxides has focused primarily on bulk parameters to determine if defects are present within the material. X-ray diffraction (XRD) studies of the strong dependence of the film lattice parameters on stoichiometry have traditionally been used to evaluate the growth window [2,8–11]. However, XRD and transport measurements are primarily sensitive to the bulk properties of a film and surface studies of hMBE films have been quite limited. Understanding the film surface is critical for the engineering of emergent properties across dissimilar interfaces including polar/non-polar heterostructures.

To date, in most heterostructures grown by hMBE only the $A$-site cation has been varied, while chemical continuity was maintained across the $B$-site, such as in $BaSnO_3/SrSnO_3$ or $ATiO_3/SrTiO_3$ multilayer films [21–23]. However, in multilayer structures combining dissimilar $A$-site and $B$-site cations, control of the interfacial termination is critical to designing functional properties[3,24,25]. The surface of hybrid-grown STO films is at present poorly understood, and growth of complex heterostructures is difficult to achieve in a controlled and predictive manner. Furthermore, in spite of decades of studies, the stability of various STO surface terminations and reconstructions is still a matter of great debate[26–29]. In the case of hMBE growth, these studies are confounded by the growth environment, where the residue of the metal-organic precursor may introduce carbon and hydrogen as well as the transition metal cation. In most STO hMBE film growth, *in situ* reflection high energy electron diffraction (RHEED) has been applied to monitor the growth in real time domain, though *in-vacuo* X-ray photoelectron spectroscopy (XPS) studies have never been reported. RHEED measurements have been important for quantifying the surface reconstructions to establish a growth window, with reports that a c(4×4) surface reconstruction is indicative of the best film stoichiometry as determined through XRD and electron mobility measurements[11]. However, the actual surface termination and chemistry of hybrid-grown STO films has never been reported, which is the subject of this work.

Here we focus on characterization of the surface stoichiometry of the hMBE grown film using *in vacuo* XPS[30] and quantify the critical growth window. Measurement of the surface termination and reconstruction of hMBE grown oxides helps to explain the relationship between the surface chemical reactions and the growth window, which has previously been reported for the bulk material and was established with XRD. From previous work, it is clear that the chemical reactions between the surface of the STO film, the TTIP precursor, and the atomic and molecular oxygen supplied during growth are quite complex. Brahlek et al. hypothesized that TTIP would preferentially decompose on the portion of the STO surface that has an SrO termination to produce stoichiometric growth, suggesting that the decomposition is more efficiently catalyzed on this termination[9]. *In-vacuo* characterization by XPS provides a route to understand the surface chemistry during growth and address these questions. Here we examine the surface termination of films in relation to the stoichiometric growth window that is observed in XRD. We also examine the role of water, carbon, and hydrogen adsorption due to residual hydrocarbons in the growth chamber after TTIP decomposition[12,27]. We examine the surface of STO films grown by hMBE using *in situ* and *ex situ* surface techniques, including RHEED, XPS, scanning tunneling microscopy (STM), and low-energy electron diffraction and microscopy (LEED/LEEM) to understand the stable surface under the unusual growth conditions present during hMBE. We show that the presence of surface SrO is crucial to the growth of stoichiometric STO and subject to a complex surface chemistry which is not yet well understood.

## II. EXPERIMENTAL

STO thin films were grown on air annealed $10\times10\times0.5$ mm$^3$ (001) STO single crystal substrates (MTI Crystal) in a Mantis MBE reactor with a baseline pressure of ~$10^{-9}$ Torr. Air annealing of the STO substrates in a tube furnace at 1000 °C maintains a mixed surface termination of $TiO_2$ and SrO because no buffered oxide etching is employed[31,32]. The mixed termination was confirmed through angle-resolved XPS measurements shown in the supplemental information (Supplemental Figure S7).[33] A mixed termination was chosen to avoid biasing the growth to a particular termination so as to probe the equilibrium surface chemistry for hybrid MBE. An additional sample was prepared on an annealed STO substrate by depositing one monolayer of SrO to produce a nominally SrO-terminated STO sample for XPS calibration. A $TiO_2$-terminated STO substrate was also prepared as a reference sample using a well-established buffered oxide etchant annealing recipe[31].

The substrates were ultrasonically cleaned in acetone and isopropanol respectively and dried with dry nitrogen gas. All substrates were cleaned in oxygen plasma in the MBE growth chamber by ramping to the 1000 °C growth temperature over ~1 hour as measured by thermocouple[10,11]. We estimate that the setpoint is ~150-200 °C higher than substrate surface temperature due to the absence of backside substrate metallization, resulting in an STO surface temperature between 800 and 850 °C. Growth and cleaning were performed in a background pressure of ~$3\times10^{-6}$ Torr $O_2$ with an RF plasma source driven at 300 W to supply O atoms to increase oxygen chemical potential at the substrate surface. Strontium (99.99%, Sigma-Aldrich, USA) was supplied through a low temperature effusion cell. The flux of SrO was calibrated using the QCM under oxygen environment with a chamber pressure of $3\times10^{-6}$ Torr. Titanium was supplied through a gas source using a metalorganic precursor TTIP (99.999%, Sigma-Aldrich, USA) from a bubbler that was connected through the gas inlet system to the growth chamber [10]. No carrier gas was used. The growth chamber shroud walls were maintained at -30 °C via a closed loop chiller and low temperature fluid (Syltherm XLT, Dow Chemical) to reduce the background water vapor pressure from the dissociated TTIP molecules. However, the chamber pressure measured by cold cathode gauge generally increases and reaches up to ~$5\times10^{-6}$ Torr for the first 5-10 minutes during growth before stabilizing due to the dissociated and unreacted TTIP injected into the system. The plasma was left on while cooling the film to below 300 °C.

A series of seven samples was grown at differing TTIP:Sr flux ratios where the Sr flux was held fixed and the TTIP gas inlet line pressure was varied. Samples are coded numerically throughout this work by the ratio of the pressure of the TTIP in the gas injection line in mTorr to the flux of SrO measured by quartz crystal microbalance (QCM) in Å/sec. While the units of these numbers cannot be compared to the literature, they serve as a proxy for comparison to reports made by other groups using beam flux pressures[10]. Drift in the Sr flux during the course of the growth is the primary source of uncertainty in TTIP:Sr flux ratios and is assumed to produce a 2% error bar in the ratios, which is consistent with other reports of cell drift[34].

*In situ* RHEED (Staib Instrument) was used to monitor the growth process and the quality of the film. RHEED intensity oscillations were used to estimate the growth rate, which was corroborated by measuring the spacing of the interference fringes in XRD scans to calculate the film thickness and dividing by the overall growth time. Growth rates were ~4.5 Å/min. Videos of the RHEED patterns were recorded and analyzed using k-means clustering and principal component analysis (PCA) codes developed at Auburn and inspired by previous work by Vasudevan et al.[35]. PCA compresses the data while retaining the most statistically relevant features, producing a new dataset of orthogonal principle components and time-dependent loadings. The reduced data set is grouped into time clusters using k-means clustering, an unsupervised machine learning technique that groups the frames based on their statistical distance to the mean image of the cluster. This analysis captures any changes or transitions in the RHEED pattern at the time of occurrence in the growth. More information about this technique is described elsewhere[36].

Post-growth the samples were transferred from the MBE reactor to the PHI 5400 XPS (monochromatic Al Kα X-ray source) system through ultra-high vacuum (UHV) transfer line (~$1\times10^{-8}$ Torr)[30]. Core level peaks were measured with a pass energy of 17.9 eV. An electron neutralizer gun was applied to compensate charging of the insulating samples. The surface stoichiometry of all grown samples was characterized by analyzing core level XPS spectra measured with base pressure of ~$1\times10^{-9}$ Torr. The total time from cooldown to the beginning of XPS measurements Analysis of the XPS data was performed using CasaXPS, with a Shirley background subtraction and Gaussian-Lorentzian (Voigt) peak shape. To properly align all core level peaks, the O 1s peak was aligned to place the lowest binding energy peak at 530.0 eV, which corresponds to oxygen bonded to metal cations in the STO film. Measurements were made at 45° and 70° electron takeoff angles to vary the degree of surface sensitivity, with 45° providing a more bulk-sensitive probe of stoichiometry and 70° a more surface-sensitive. We assume in our analysis that electron forward focusing effects, commonly referred to as X-ray photoelectron diffraction, are negligible between samples[37]. A Malvern Panalytical X'Pert³ Diffractometer equipped with a 4-circle goniometer and using Cu $K\alpha_1$ radiation line isolated with double bounce Ge (111) monochromator was used for $2\theta$-$\omega$ scans on the (002) reflection of the STO samples.

Further LEED/LEEM measurements were carried out in the XPEEM/LEEM endstation of the Electron Spectro-Microscopy (ESM, 21-ID-2) beamline at the National Synchrotron Light Source II [38] on an STO sample grown on a Nb-doped STO substrate. The sample was degassed at 120 °C with a pressure of $6\times10^{-10}$ Torr for 1 hour. The sample was then annealed in a preparation chamber at a pressure of $2\times10^{-6}$ Torr of $O_2$ at 500 °C for 30 mins and cooled down in the same pressure of $O_2$ to room temperature to remove surface contamination. In this system, the electron-beam spot size in the LEED mode is 1.5 μm in diameter and the spatial resolution in the LEEM mode is < 10 nm[39]. The base pressure of the analysis chamber is $2\times10^{-10}$ Torr.

STM was performed on the STO sample grown on a Nb-doped STO substrate with a base pressure of $3.0\times10^{-10}$ Torr using an Omicron Nanotechnology Variable Temperature Scanning Probe Microscopy (VT-SPM) system. Imaging was conducted with an electrochemically etched tungsten tip at ambient temperature, and the topography images were analyzed with the open-source software Gwyddion [40]. The images have been leveled, and the color range was adjusted for the purpose of illustration. The sample was heated to 450°C with $3\times10^{-8}$ Torr of $O_2$ for 3 hours prior to imaging to remove ambient contamination. Imaging condition, unless stated otherwise, were bias voltage $V_{bias}$=2.0V, and tunneling current $I_t$=0.1 nA which yielded the best imaging conditions. The residual roughness on the terraces can be attributed to air exposure of the sample; more aggressive in-situ treatment was avoided to conserve the terrace structure formed after growth. Higher resolution images of the round features on the terraces remained elusive but the morphology on the terraces was observed consistently over the entire sample surface.

## III. RESULTS AND DISCUSSION

Figure 1 shows the RHEED patterns after growth and cool-down for five of the seven samples grown on undoped STO substrates along <110> and <100> azimuth. The RHEED patterns show strong diffraction peaks and additional intermediate peaks due to surface reconstructions in films with low TTIP:Sr flux (ratios of 740, 780, and 856) compared to films with a higher flux ratio. Images for samples 758 and 798 are shown in the Supplemental Information[33]. The strong Kikuchi lines reflect the high crystal quality of the film. Faint surface reconstructions are observed for samples 740, 780, and 856 along (100) and/or (110) but specific determination of the nature of the reconstruction is challenging. While sample 880 shows clear spotty features due to higher defect concentrations associated with off-stoichiometry.

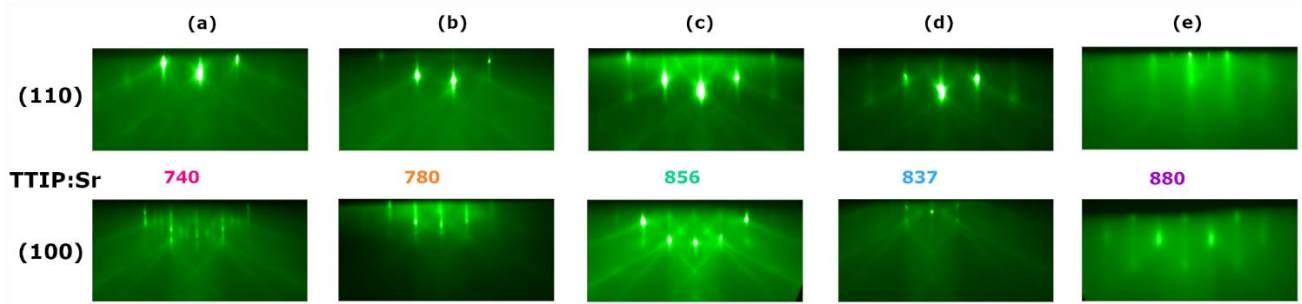

*Figure 1: RHEED images (a)-(e) along 110 and 100 planes with increasing Ti:Sr peak area ratios as measured by XPS. Color-coded numbers match the samples to those in Figure 2. Calibrated flux ratios for each sample are shown in the center. Samples are ordered left to right based on the measured Ti 2p:Sr 3d peak area ratio shown in Figure 2(a).*

XPS measurements immediately after growth focused on quantifying Ti:Sr peak area ratios to determine surface and bulk film stoichiometry. Figure 2(a) shows the normalized area ratio of Ti 2p to the Sr 3d core level as a function of TTIP:Sr flux at photoelectrons takeoff angles of 45° and 70°. The dashed grey, red and purple lines represent the *B*-site to *A*-site core level area ratio for an air annealed STO substrate, air annealed Nb-doped STO substrate, and STO substrate after deposition of one monolayer of SrO, respectively. These references were helpful to maintain the same nominal stoichiometry and cation ratio for the STO film grown on Nb:STO substrate. With increasing TTIP:Sr flux the area ratio also increases, except for the sample with a TTIP:Sr ratio of 856. Deviation in XPS peak area ratios for the 856 sample away from the monotonic increase shown in the other samples is most likely attributable to drift or slight miscalibration in the Sr source flux. As the surface sensitivity of the XPS measurements increases with electron emission angle, the changes in area ratio can be used to infer the primary surface termination of the sample. In fact, simply by changing the STO termination from SrO to $TiO_2$, one can expect to see a significant change in the Ti 2p:Sr 3d peak area ratio of approximately 20% (see Supplemental Information Section 1 for details[33]).

In addition to studies of the Sr:Ti cation ratios, possible carbon contamination has also been resolved in the STO films through XPS. This carbon would be present from dissociation of the metalorganic precursor. As the TTIP precursor contains large amounts of carbon, there have been long-standing questions as to the presence of carbon in the bulk of the film or on the surface[12]. In our core level XPS scan (see Supplemental Information Figure S1[33]), we found the C 1s peak was negligibly small in comparison with the nearby Sr 3p peaks, which could be from the adsorption of carbon during the time required to transfer the sample and perform XPS measurements. The level of carbon contamination is consistent with what would be expected from adsorption, but we cannot rule out some residual carbon from the precursor. Together with previous measurements[12], however, our results suggests that hMBE samples grown with activated oxygen plasma do not have appreciable carbon contamination. Whether this is also the case for samples grown in molecular oxygen will be determined in future work.

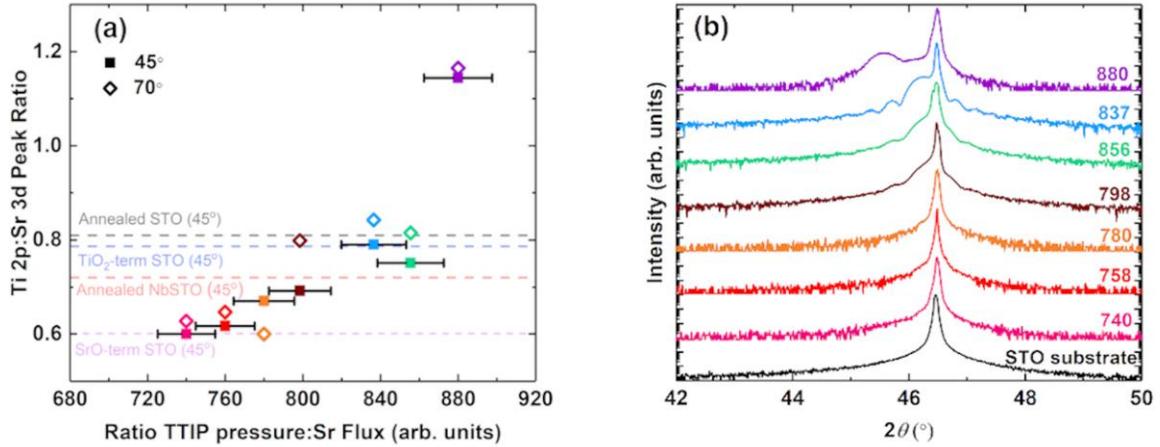

*Figure 2: (a) Normalized area ratio of Ti 2p$_{3/2}$ to Sr 3d$_{5/2}$ as a function of TTIP pressure to Sr flux ratio at 45° and 70° electron emission angle in XPS. Error bars for the peak area ratios are smaller than the graphical data points. (b) Intensity as a function of 2θ for XRD scans on (002) Bragg peak.*

In Figure 2(b), XRD scans are shown for all seven films along with an air-annealed plain STO substrate (black). A single XRD peak for a homoepitaxial film results from out-of-plane lattice matching between the substrate and film and is indicative of a stoichiometric sample with a pristine film-substrate interface with no contamination from atmospheric elements such as carbon[19]. Off-stoichiometric samples produce a secondary film peak at smaller values of 2θ consistent with a larger out-of-plane lattice parameter. The 740, 758, and 780 samples are stoichiometric whereas the 798 and 856 are slightly off-stoichiometric. The 837 and 880 samples have clear secondary peaks indicating significant off-stoichiometry in the films.

In comparison to the XRD measurements, XPS is highly surface sensitive. The more bulk sensitive 45° measurements from XPS for stoichiometric samples (740, 758 and 780) reflect a smaller than expected Ti 2*p*:Sr 3*d* ratio when compared to the single crystal references, with ratios between 0.6 and 0.7. At the more surface sensitive 70° emission angle, we observe that the Ti 2*p*:Sr 3*d* area ratio of the 780 sample decreases, indicating a majority SrO termination. Meanwhile, samples 740 and 758 change only slightly, suggesting a mixed termination with greater concentrations of surface SrO than either of the single crystal references. Comparisons of Ti 2p peak fits for stoichiometric and off-stoichiometric samples (see Figure S8) confirm that there is no difference in Ti valence between samples, indicating that the off-stoichiometry is not attributable to differences in oxygen vacancy concentrations between samples. All samples show exclusively $Ti^{4+}$ peaks in the Ti 2*p* spectra. Previous work has indicated that changes to oxygen stoichiometry produce only 0.001-0.002 Å change in lattice parameter, which is barely detectable even in bulk samples[41] and thus not likely to cause the diffraction peaks we measure.

Given the observed XPS peak area ratio from Figure 2(a), it seems likely that the off-stoichiometric samples (798, 837, 856, and 880) are Ti rich and contain Sr vacancies that expand the lattice constant of the film. Conversely, samples with Ti 2*p*:Sr 3*d* area ratios less than those of the single crystal substrates (samples 740, 758, and 780) were found to be stoichiometric as measured by XRD. The similarity of sample 798 to the Nb:STO reference suggests that it is majority $TiO_2$ terminated. The $TiO_2$ termination, greater peak area ratio, and greater TTIP flux in comparison to the three stoichiometric samples indicate that the sample is Ti rich. This conclusion can be attributed to some excess Ti floating towards the surface, as has been reported by others for off-stoichiometric STO growth [30,42–44]. The nearly monotonic increase in area ratio suggests that the other three samples are also Ti rich.

This somewhat surprising result suggests that more Sr is present on the film surface relative to the annealed reference substrates. By comparing the peak area ratio to the SrO-terminated STO substrate, we see that the peak area ratios at 45° are very close to that of SrO-terminated STO. As the XPS becomes more surface sensitive the weight of signals coming off the top SrO monolayer is significantly greater relative to the 45° measurements resulting relatively low area ratio (See Supplemental Information Figure S7[33]). However, the variation in peak area ratios between the more bulk-sensitive 45° and more surface-sensitive 70° measurements indicates that the stoichiometric film surfaces are not fully SrO terminated. Meanwhile, the angle-resolved measurements for the off-stoichiometric samples show that they do exhibit majority $TiO_2$ terminations.

With the unexpected results from the stoichiometry measurements, an understanding of the surface chemistry gives insight into the film growth process. To understand the surface chemistry of the films, the O 1s peak was further analyzed to determine the surface bonding environment. Figure 3 shows the deconvolution of the O1s core level using position constraints of 1.1 eV and 2.3 eV higher binding energy from the bulk oxygen peak for OH and surface oxygen respectively that has been applied elsewhere [45]. The two surface peaks at 1.1 and 2.3 eV have higher binding energies (531.1 eV and 532.3 eV, respectively) than the STO lattice oxygen, and are more pronounced in the 70° orientation compared to the 45° configuration due to the greater surface sensitivity provided in angle resolved XPS. Therefore, the 70° XPS measurement better quantifies the adsorbed water, hydroxyl, and undercoordinated oxygen on the surface of the film. A more pronounced 532.3 eV surface peak is observed for samples 740 and 780 (Figure 3(a-b)). The absence of the 532.3 eV surface component for films with higher TTIP:Sr flux ratio indicates that these films are primarily Ti- terminated or Ti-rich (Figure 3(c-e)). The 532.3 eV surface peak was also observed in the SrO-terminated sample, as shown in Supplemental Figure S6[33].

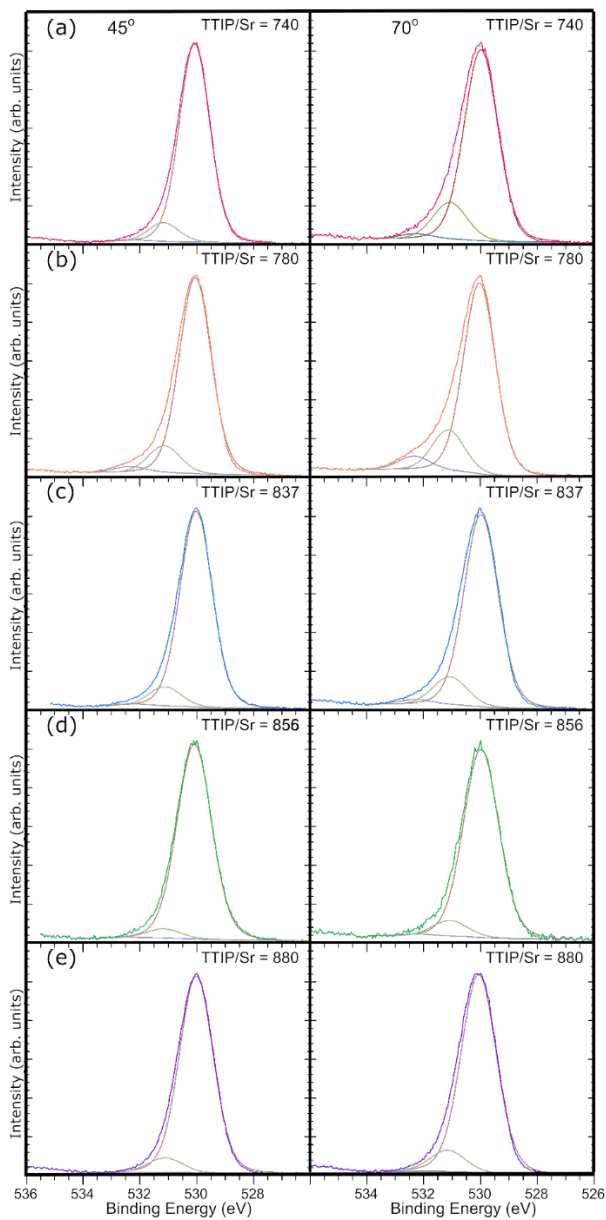

*Figure 3: O 1s core level deconvolution with electron emission angle 45° and 70° through (a)-(e). Peaks at 530.0 eV, 531.1 eV, and 532.3 eV correspond to lattice oxygen, hydroxyl (OH), and the surface oxygen feature, respectively.*

Given that sample 740 has been shown to be stoichiometric and appears to exhibit a partial SrO termination, it was chosen for further analysis of the RHEED videos to understand the growth process. PCA of the RHEED video for Sample 740 is displayed in Figure 4[36]. By studying the time dependence of the principal components, we can glean the typical information inferred from RHEED oscillations with additional understanding of what components of the surface morphology vary with time. The first two principal components largely contain features that do not appear to correlate with surface evolution during growth. These peaks may originate from a combination of background noise due to cryopump vibrations, diffraction off the bulk of the STO substrate, and the centroid of the zeroth order RHEED streak, which would all be relatively uncorrelated with the film surface evolution. Conversely, oscillations with a frequency that corresponds to a growth rate of 0.45 nm/min are apparent in components 3 through 5. The features of components 3 through 5 are consistent with 2D growth modes observed previously by Vasudevan *et al*[35]. These oscillations damp out around 1100-1300 seconds and the RHEED pattern remains uniform for the remainder of the growth. This transition is consistent with the disappearance of conventional RHEED oscillations reported by Kajdos and Stemmer for stoichiometric films, which was attributed to a transition to step-flow growth [11]. However, similar results could likely be observed for any surface that has reached an equilibrium termination and maintains a uniform surface roughness. Analysis of the RHEED patterns through K-means clustering (see Supplemental Information Figure S2 for more details[33]) further supports this interpretation, as the pattern for Sample 740 transitions from an initial pattern (or "cluster") over the first ~1100 seconds to a final cluster that has minimal time variation for the remainder of the growth. Collectively, these results indicate that the film surface transforms over the first ~1100 seconds (~25 unit cells) from the poorly defined surface present on the air-annealed substrate to an equilibrium surface that is largely unchanged for the remainder of the growth.

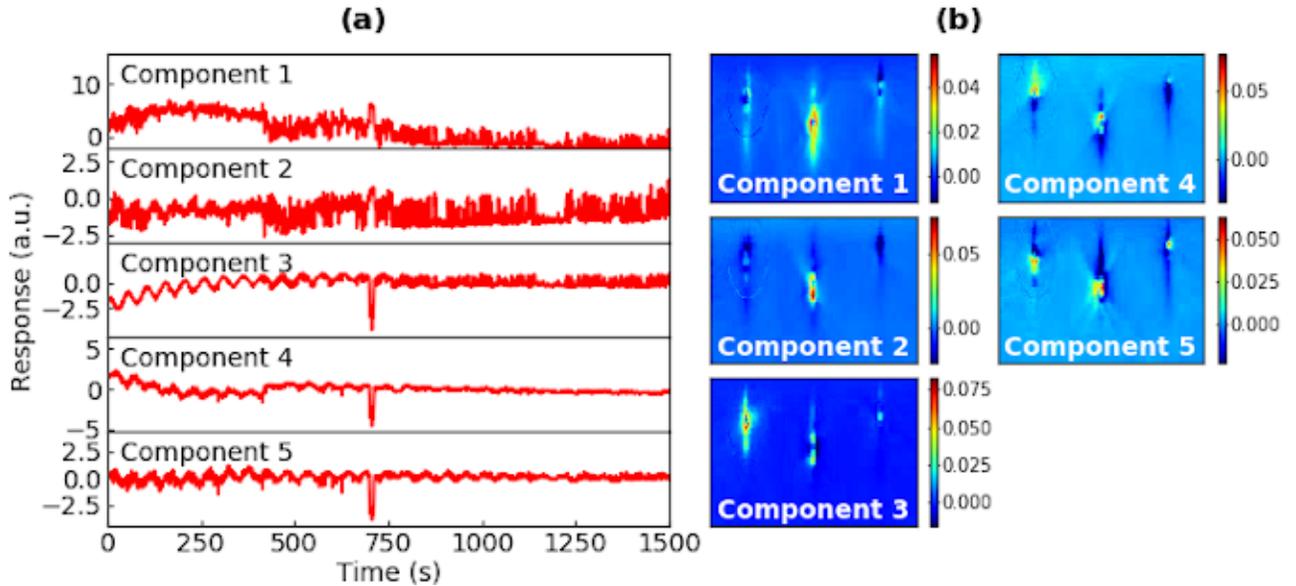

*Figure 4: Time-dependent eigenvalues (left) and eigenvectors (right) for sample (a) with a constant TTIP/Sr flux ratio of 740.*

To summarize the results from the homoepitaxial samples presented thus far, we find that the XRD curves in Figure 2 verify that films with greater than expected Sr concentrations at or near the surface are stoichiometric. Meanwhile those that exhibit Ti:Sr peak area ratios comparable to or greater than what is seen in the annealed STO substrates are non-stoichiometric based on XRD measurements and are presumably Ti-rich. The observed 532.3 eV O 1*s* peak for partially SrO terminated films compared to the Ti-rich samples is indicative of undercoordinated oxygen, oxygen vacancies, hydroxyl formation, or other variations in oxygen chemistry at the STO film surface. PCA of the RHEED videos for a stoichiometric film (sample 740) implies uniform growth of the films after ~1100 secs with no changes in surface termination or roughness. This is consistent with the growth of stoichiometric samples reported previously [10,11,14] and indicates that the film surface is in chemical equilibrium with the Sr adatom flux and TTIP and oxygen vapor. Given the relative increase in surface Sr that we observe for stoichiometric Samples 740, 758, and 780, this suggests that the equilibrium surface for STO films grown by hybrid MBE within the growth window has a partial SrO termination and greater Sr concentrations at the surface than air-annealed single crystal substrates.

To further understand the apparent SrO surface termination of the samples studied thus far, an additional sample was grown with the same nominal stoichiometry and cation ratio (~0.6) from the XPS analysis as sample 740 on Nb-doped STO (see Supplemental Information Figures S3 and S4 for RHEED and XPS data respectively[33]). LEED, LEEM and STM measurements were performed for the STO film grown on Nb:STO to visualize the surface termination, reconstruction and atomic-scale morphology of the film. Figure 5 shows LEED and LEEM data for this sample. The LEED image taken with 45 eV electron energy (Figure 5(b)) clearly shows the presence of extra diffraction spots between primary spots indicating a surface reconstruction on the film surface. A c(2×2) reconstruction pattern is observed in Figure 5(b), as shown by the yellow circles. A separate faint 2×2 reconstruction can also be seen at some energies (not shown), but was not sufficiently intense for isolation in LEEM studies. A bright field LEEM (BF-LEEM) image at 20 eV and dark field LEEM (DF-LEEM) image with c(2×2) at 22 eV (Figure 5(c-d)) show that the reconstruction is not uniform across the whole surface. As the c(2×2) spot was intense enough, the DF-LEEM image was reproduced using this spot and BF-LEEM image was reproduced using the specular reflection. These results are consistent with a mixed surface termination, as has been observed BF-LEEM for oxide surfaces prepared by annealing[29,32]. The dark contrast in the BF-LEEM image has been attributed to differences in the surface potential or work function for a SrO termination in comparison to a TiO$_2$ termination on the surface. The DF-LEEM further confirms that the dark regions in the BF-LEEM image correspond to regions with a c(2×2) surface termination, which is commonly associated with a Sr-rich or Sr-terminated surface[11].

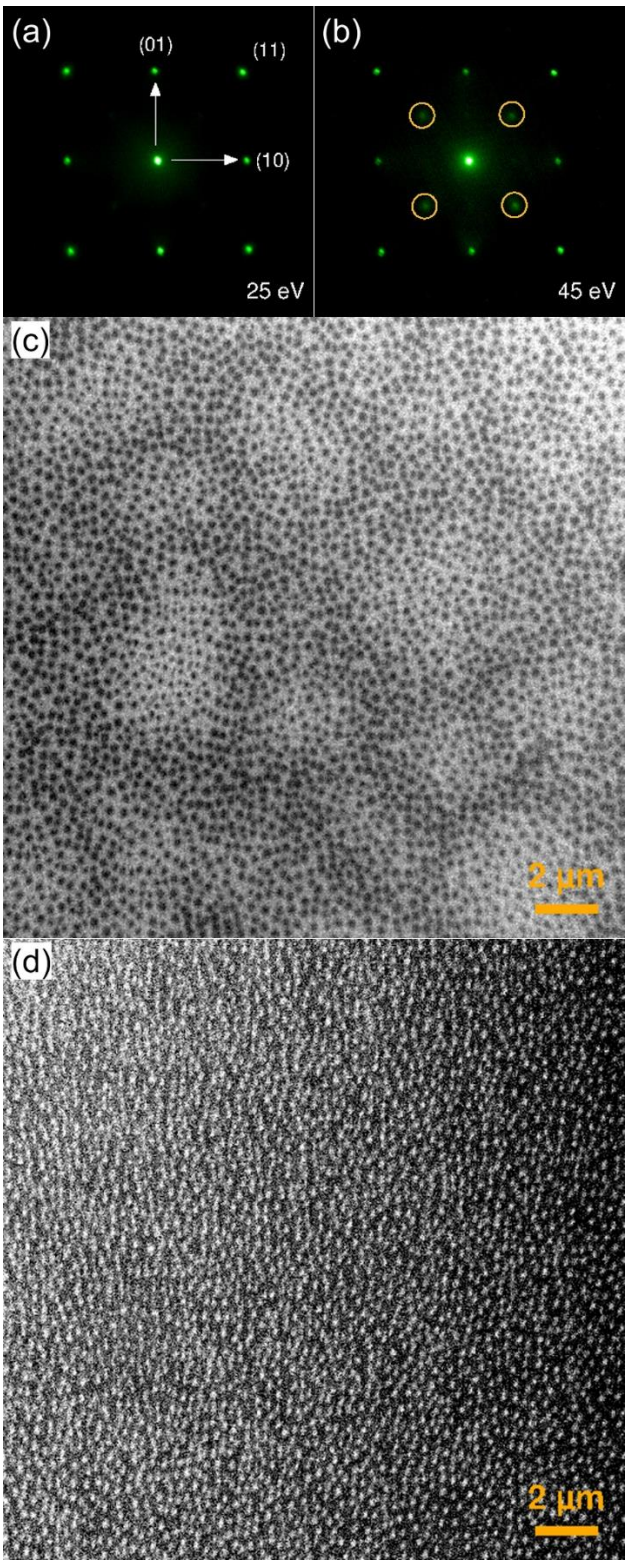

*Figure 5: Low energy electron diffraction (LEED) at (a) 25 eV and (b) 40 eV electron energies, yellow circles highlight the c(2×2) reconstructions; (c) Bright field low energy*

*electron microscopy (LEEM) image; and (d) Dark field LEEM image highlighting the regions of c(2×2) reconstructions.*

To examine the island structures observed through LEED and LEEM measurements, the sample was also analyzed through STM imaging experiments. Figure 6 shows a characteristic wedding-cake type structure observed across the entire sample with terrace widths of 40 to 70 nm on stacks with well-defined steps. Atomic force microscopy measurements on the homoepitaxial series of samples show qualitative agreement with the morphology observed in this sample (see Supplemental Information Figure S9 [33]). A wedding cake structure is formed when the Ehrlich-Schwöbel (ES) barrier, which limits diffusion across the step edge, cannot be overcome. The adsorbates which ultimately form the growing oxide, are trapped on the top layer and reflected from the step edge. The adsorbate diffusion length on the surface is correlated to the width of the rim – the distance between the terrace step edges[46]. Ultimately, once the adsorbate concentration is sufficiently high, a new layer will nucleate, albeit with a diminished radius which leads to the stacked, wedding cake like structure. The apparent height of the step edges is about ~4 Å which corresponds to a unit cell step height of STO with lattice constant 3.905 Å. In contrast to the strong step-edge anisotropy often observed in metal-on-metal epitaxy with a significant ES barrier [47], the oxide islands do not exhibit a pronounced edge faceting or anisotropy. STM studies of (La,Ca)MnO$_3$ perovskite oxide films have also demonstrated the wedding cake structure due to the ES barrier in these materials [48–50]. These studies examined transitions between mixed terminations and uniform *B*-site termination as a function of film thickness and noted the presence of a c(2×2) reconstruction for the *A*-site terminated regions, as we also observe in the LEED/LEEM studies. Given our results, it seems apparent that the homoepitaxial growth mode for hybrid-grown STO films within the stoichiometric growth window is not exclusively layer-by-layer or step-flow.

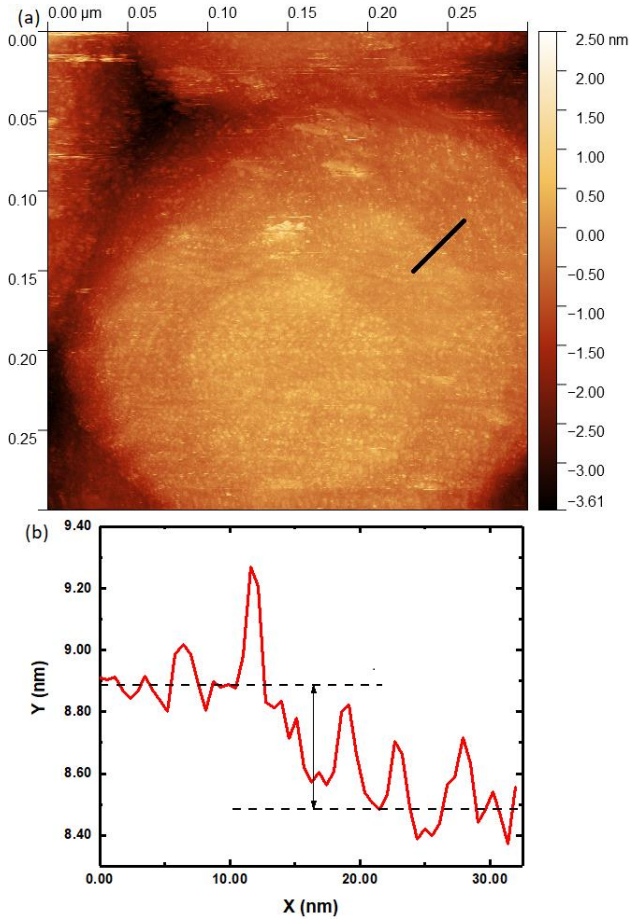

*Figure 6: (a) Scanning tunneling microscopy (STM) image with (b) line profile shown from black line in (a). Dashed lines represent a 4.0 Å step consistent with the 3.905 Å SrTiO$_3$ unit cell height. Image size 300×300 nm².*

The surface studies that we report here raise questions as to the nature of the chemical interaction during the hMBE growth process. We have shown that samples with partial SrO termination (samples 740, 758, and 780) are stoichiometric in XRD. Conversely, the XRD results demonstrate samples with surface compositions analogous to that of an air-annealed substrate are off-stoichiometric. Based on angle-resolved XPS measurements, these off-stoichiometric samples showed a preference for majority $TiO_2$ surface terminations. Surface reconstructions are observed for stoichiometric samples, although we have not observed the c(4×4) reconstruction that others have reported[11]. Instead, a surface with partial SrO termination and evidence of a c(2×2) surface reconstruction in some cases aligns with the ideal stoichiometry as measured by XRD. These *in situ* observations are confirmed by the LEEM, LEED, and STM measurements. While a surface reconstruction could not be resolved by STM due to atmospheric exposure, the rounded nature of the wedding cake terraces suggests that the terraces are the source of the c(2×2) reconstruction observed in the LEEM images of Figure 5(c-d). Thus it is likely that the wedding cake terraces have a majority SrO termination, which gives rise to the significant ES barrier.

Our results suggest that in order to stabilize the adsorption-controlled growth regime, a significant portion of the surface must possess an SrO termination to catalyze the decomposition of TTIP in stoichiometric ratios. Brahlek *et al*[9] have previously hypothesized that the TTIP sticking coefficient is greater on an SrO termination than on a $TiO_2$ termination, which agrees with the results we show here. Recent theoretical studies have predicted a stable c(4×4) SrO-terminated STO surface with only half of the surface A sites occupied, which would qualitatively agree with our results and merits further investigation[51]. The existence of the extra XPS peak in the O 1*s* spectrum is associated with undercoordinated oxygen, oxygen vacancies, adsorbed water, or bonded OH on the surface for SrO terminations, which is not observed for $TiO_2$ terminated films. The presence of adsorbed water or oxygen vacancies will depend significantly on the surface termination and the ambient environment[26,27]. Given that water is a natural byproduct of the TTIP dissociation, the role of water vapor near the film surface may play a role in the chemistry of the adsorption-controlled growth. In the future, *in situ* STM, LEEM, and LEED studies of hMBE grown STO films would be extremely valuable to understand the surface structure in greater detail. However, to our knowledge, there is not currently a growth system configured for such studies. Further development of *in situ* surface characterization capabilities to complement hMBE growth would be extremely valuable.

## IV. SUMMARY AND CONCLUSIONS

In conclusion, we have demonstrated hMBE synthesis of homoepitaxial STO films with *in vacuo* XPS studies for the first time. We show that carbon contamination on the surface is negligible. We find that films exhibiting a partial SrO termination are needed to produce a stoichiometric film as measured in XRD, while $TiO_2$ terminated films are Ti rich. These results are confirmed via XPS through core level Ti 2p to Sr 3d peak area ratios and the observation of O 1s surface peaks. Analysis of the RHEED pattern through PCA and K-means clustering indicate that the stoichiometric films transition to a step-flow growth regime after 20-25 unit cells. RHEED, LEEM, LEED, and STM studies verify that surface reconstructions are associated with ideal film stoichiometry, though other complementary surface studies are needed to better probe the reconstructions and surface terminations. Understanding the role that different precursor chemistries and growth environments play on the synthesis process will be critical to the continued evolution of the hybrid MBE technique. Future *in situ* studies of complex oxide films grown by hybrid MBE are needed to help better explain the nature of the growth process for different precursor chemistries.

# ACKNOWLEDGMENTS

We acknowledge Z. Dai for technical assistance. ST and RBC gratefully acknowledge support from the Air Force Office of Scientific Research under award number FA9550-20-1-0034. ST, SRP, and WJ also acknowledge support from the Auburn University Department of Physics. JML and MB acknowledge support from by the U.S. Department of Energy (DOE), Office of Science, Basic Energy Sciences, Materials Sciences and Engineering Division. This research used resources of the Center for Functional Nanomaterials and National Synchrotron Light Source II, which are U.S. DOE Office of Science Facilities, at Brookhaven National Laboratory under Contract No. DE-SC0012704. DJ and PR acknowlegde the support by support from the National Science Foundation (NSF) (Award No. DMR-2004326) by the Division of Materials Research – Metals and Metallic Nanostructures.

# DATA AVAILABILITY

The data that supports the findings of this study are available within the article and its supplementary material. Raw data that support the findings of this study are available from the corresponding author upon reasonable request.

# REFERENCES

[1] A. Ohtomo and H.Y. Hwang, Nature **427**, 423 (2004).
[2] S.A. Chambers, Surface Science **605**, 1133 (2011).
[3] S.R. Spurgeon, P.V. Sushko, S.A. Chambers, and R.B. Comes, Phys. Rev. Materials **1**, 063401 (2017).
[4] M. Nakamura, F. Kagawa, T. Tanigaki, H.S. Park, T. Matsuda, D. Shindo, Y. Tokura, and M. Kawasaki, Phys. Rev. Lett. **116**, 156801 (2016).
[5] R. Comes and S. Chambers, Physical Review Letters **117**, 226802 (2016).
[6] K. Nakamura, H. Mashiko, K. Yoshimatsu, and A. Ohtomo, Applied Physics Letters **108**, 211605 (2016).

[7] J.L. MacManus-Driscoll, M.P. Wells, C. Yun, J.-W. Lee, C.-B. Eom, and D.G. Schlom, APL Materials **8**, 040904 (2020).

[8] J.H. Haeni, C.D. Theis, and D.G. Schlom, Journal of Electroceramics **4**, 385 (2000).

[9] M. Brahlek, A.S. Gupta, J. Lapano, J. Roth, H.-T. Zhang, L. Zhang, R. Haislmaier, and R. Engel-Herbert, Adv. Funct. Mater. **28**, 1702772 (2018).

[10] B. Jalan, R. Engel-Herbert, N.J. Wright, and S. Stemmer, Journal of Vacuum Science and Techology A **27**, 461 (2009).

[11] A.P. Kajdos and S. Stemmer, Applied Physics Letters **105**, 191901 (2014).

[12] B. Jalan, J. Cagnon, T.E. Mates, and S. Stemmer, Journal of Vacuum Science & Technology A: Vacuum, Surfaces, and Films **27**, 1365 (2009).

[13] J. Son, P. Moetakef, B. Jalan, O. Bierwagen, N.J. Wright, R. Engel-Herbert, and S. Stemmer, Nat Mater **9**, 482 (2010).

[14] B. Jalan, P. Moetakef, and S. Stemmer, Applied Physics Letters **95**, 032906 (2009).

[15] M. Brahlek, L. Zhang, C. Eaton, H.-T. Zhang, and R. Engel-Herbert, Appl. Phys. Lett. **107**, 143108 (2015).

[16] A.P. Kajdos, D.G. Ouellette, T.A. Cain, and S. Stemmer, Applied Physics Letters **103**, 082120 (2013).

[17] B. Jalan, S.J. Allen, G.E. Beltz, P. Moetakef, and S. Stemmer, Appl. Phys. Lett. **98**, 132102 (2011).

[18] A. Prakash, P. Xu, X. Wu, G. Haugstad, X. Wang, and B. Jalan, J. Mater. Chem. C **5**, 5730 (2017).

[19] J.M. LeBeau, R. Engel-Herbert, B. Jalan, J. Cagnon, P. Moetakef, S. Stemmer, and G.B. Stephenson, Applied Physics Letters **95**, 142905 (2009).

[20] J. Roth, E. Arriaga, M. Brahlek, J. Lapano, and R. Engel-Herbert, Journal of Vacuum Science & Technology A **36**, 020601 (2017).

[21] L.R. Thoutam, J. Yue, P. Xu, and B. Jalan, Phys. Rev. Materials **3**, 065006 (2019).

[22] H. Yun, J. Held, A. Prakash, T. Wang, B. Jalan, and K.A. Mkhoyan, Microscopy and Microanalysis **25**, 2110 (2019).

[23] P. Moetakef, T.A. Cain, D.G. Ouellette, J.Y. Zhang, D.O. Klenov, A. Janotti, C.G.V. de Walle, S. Rajan, S.J. Allen, and S. Stemmer, Applied Physics Letters **99**, 232116 (2011).

[24] S.-C. Lin, C.-T. Kuo, R.B. Comes, J.E. Rault, J.-P. Rueff, S. Nemšák, A. Taleb, J.B. Kortright, J. Meyer-Ilse, E. Gullikson, P.V. Sushko, S.R. Spurgeon, M. Gehlmann, M.E. Bowden, L. Plucinski, S.A. Chambers, and C.S. Fadley, Phys. Rev. B **98**, 165124 (2018).

[25] N. Nakagawa, H.Y. Hwang, and D.A. Muller, Nat Mater **5**, 204 (2006).

[26] S.A. Chambers and P.V. Sushko, Phys. Rev. Materials **3**, 125803 (2019).

[27] A.E. Becerra-Toledo, J.A. Enterkin, D.M. Kienzle, and L.D. Marks, Surface Science **606**, 791 (2012).

[28] M.B.S. Hesselberth, S.J. van der Molen, and J. Aarts, Appl. Phys. Lett. **104**, 051609 (2014).

[29] L. Aballe, S. Matencio, M. Foerster, E. Barrena, F. Sánchez, J. Fontcuberta, and C. Ocal, Chem. Mater. **27**, 6198 (2015).

[30] S. Thapa, R. Paudel, M.D. Blanchet, P.T. Gemperline, and R.B. Comes, Journal of Materials Research **36**, 26 (2021).

[31] M. Kawasaki, K. Takahashi, T. Maeda, R. Tsuchiya, M. Shinohara, O. Ishiyama, T. Yonezawa, M. Yoshimoto, and H. Koinuma, Science **266**, 1540 (1994).

[32] J. Jobst, L.M. Boers, C. Yin, J. Aarts, R.M. Tromp, and S.J. van der Molen, Ultramicroscopy **200**, 43 (2019).

[33] Placeholder for Supplemental Information (n.d.).

[34] Y.-S. Kim, N. Bansal, and S. Oh, Journal of Vacuum Science & Technology A **28**, 600 (2010).

[35] R.K. Vasudevan, A. Tselev, A.P. Baddorf, and S.V. Kalinin, ACS Nano **8**, 10899 (2014).

[36] S.R. Provence, S. Thapa, R. Paudel, T.K. Truttmann, A. Prakash, B. Jalan, and R.B. Comes, Phys. Rev. Materials **4**, 083807 (2020).

[37] S.A. Chambers, L. Wang, and D.R. Baer, Journal of Vacuum Science & Technology A **38**, 061201 (2020).

[38] R.M. Palomino, E. Stavitski, I. Waluyo, Y.K. Chen-Wiegart, M. Abeykoon, J.T. Sadowski, J.A. Rodriguez, A.I. Frenkel, and S.D. Senanayake, Synchrotron Radiation News **30**, 30 (2017).

[39] W. Jin and R.M. Osgood, Advances in Physics: X **4**, 1688187 (2019).

[40] D. Nečas and P. Klapetek, Open Physics **10**, 181 (2012).

[41] A. Spinelli, M.A. Torija, C. Liu, C. Jan, and C. Leighton, Phys. Rev. B **81**, 155110 (2010).

[42] T. Orvis, H. Kumarasubramanian, M. Surendran, S. Kutagulla, A. Cunniff, and J. Ravichandran, ACS Appl. Electron. Mater. **3**, 1422 (2021).

[43] J.H. Lee, G. Luo, I.C. Tung, S.H. Chang, Z. Luo, M. Malshe, M. Gadre, A. Bhattacharya, S.M. Nakhmanson, J.A. Eastman, H. Hong, J. Jellinek, D. Morgan, D.D. Fong, and J.W. Freeland, Nat Mater **13**, 879 (2014).

[44] Y.F. Nie, Y. Zhu, C.-H. Lee, L.F. Kourkoutis, J.A. Mundy, J. Junquera, P. Ghosez, D.J. Baek, S. Sung, X.X. Xi, K.M. Shen, D.A. Muller, and D.G. Schlom, Nat Commun **5**, (2014).

[45] K.A. Stoerzinger, R. Comes, S.R. Spurgeon, S. Thevuthasan, K. Ihm, E.J. Crumlin, and S.A. Chambers, J. Phys. Chem. Lett. **8**, 1038 (2017).

[46] J. Krug, J Stat Phys **87**, 505 (1997).

[47] M. Kalff, P. Šmilauer, G. Comsa, and T. Michely, Surface Science **426**, L447 (1999).

[48] R.K. Vasudevan, H. Dixit, A. Tselev, L. Qiao, T.L. Meyer, V.R. Cooper, A.P. Baddorf, H.N. Lee, P. Ganesh, and S.V. Kalinin, Phys. Rev. Materials **2**, 104418 (2018).

[49] A.G. Gianfrancesco, A. Tselev, A.P. Baddorf, S.V. Kalinin, and R.K. Vasudevan, Nanotechnology **26**, 455705 (2015).

[50] A. Tselev, R.K. Vasudevan, A.G. Gianfrancesco, L. Qiao, T.L. Meyer, H.N. Lee, M.D. Biegalski, A.P. Baddorf, and S.V. Kalinin, Crystal Growth & Design **16**, 2708 (2016).

[51] H.-J. Sung, Y. Mochizuki, and F. Oba, Phys. Rev. Materials **4**, 044606 (2020).

# Correlating surface stoichiometry and termination in SrTiO$_3$ films grown by hybrid molecular beam epitaxy

Suresh Thapa[1], Sydney R. Provence[1], Devin Jessup[2], Jason Lapano[3], Matthew Brahlek[3], Jerzy T. Sadowski[4], Petra Reinke[2], Wencan Jin[1], and Ryan B. Comes[1,a]

[1]Physics Department, Auburn University, Auburn, AL, 36849, USA

[2]Department of Materials Science and Engineering, University of Virginia, Charlottesville, VA 22904

[3]Materials Science and Technology Division, Oak Ridge National Laboratory, Oak Ridge, TN, 37830

[4]Center for Functional Nanomaterials, Brookhaven National Laboratory, Upton, NY, 11973

a) Electronic mail: ryan.comes@auburn.edu

## Supporting Information

### I. Ti 2p:Sr 3d peak area ratio

In order to understand how much shift in Ti 2p : Sr 3d peak area ratio could occur from experimental result if the film has completely either terminations, a simple theoretical model was created. In this model, an intensity *I(z)* of SrO and TiO$_2$ layer in a STO film was calculated by assuming complete ($A_{SrO/TiO_2} = 1$) SrO and TiO$_2$ termination in each half a unit cell STO film depth (*z*) alternately up-to 5 nm. For the x-ray incident angle ($\theta$) of 45° and wavelength ($\lambda$) 1.5 nm, all the calculated intensities are summed and evaluated the ratio of $I_{TiO_2}(z)$ to $I_{SrO}(z)$. In comparison to the STO substrate the area ratio shifts up-to ~20% depending upon the type of termination.

$$I(z) = A_{SrO/TiO_2} e^{-z/\lambda cos\theta} \quad (1)$$

### II. Carbon contamination

Figure S1 shows the core level XPS scan of a representative film with photoelectrons takeoff angle at 45°. The C 1s peak is negligibly small in comparison with the nearby Sr 3p peaks.

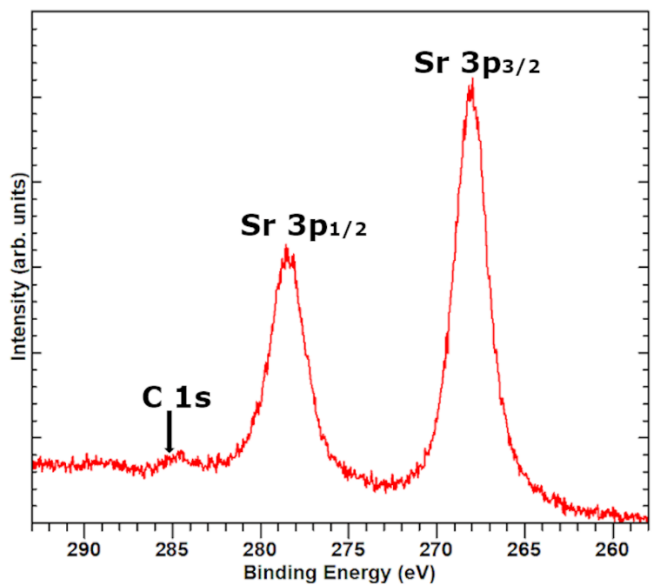

Figure S1: C1s core level with x-ray incident angle 45°.

**III. Principal Component Analysis**

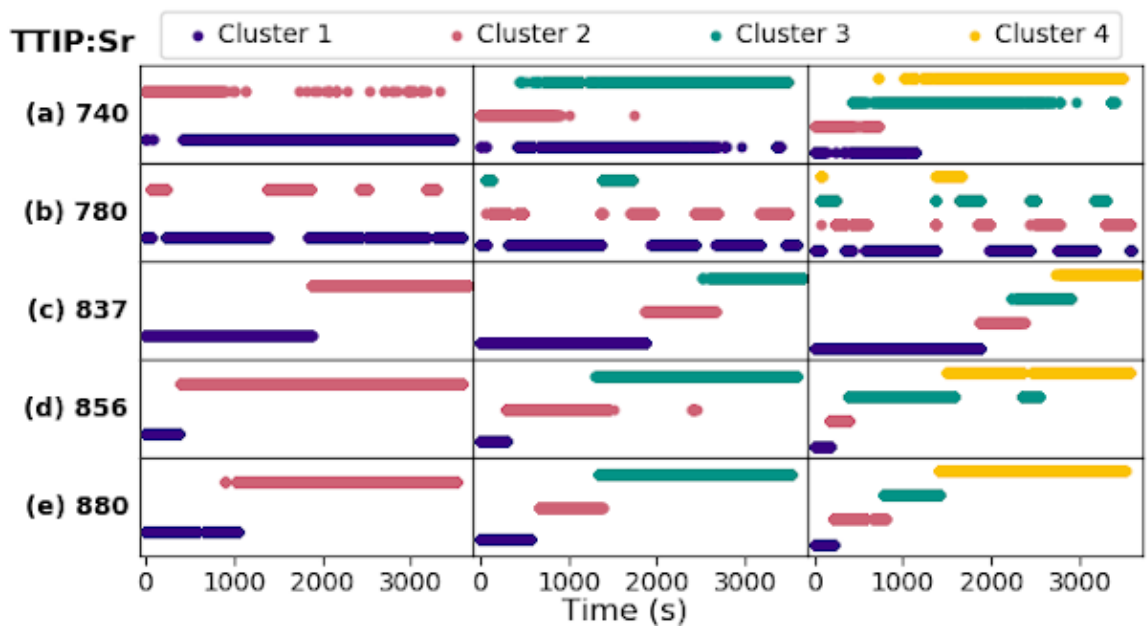

Figure S2: K-means clustering for samples (a)-(e) using 2-4 clusters.

Figure S2 shows K-means clustering for the 5 samples, with clear trends between the stoichiometric samples (740 and 780) and those that are non-stoichiometric (837, 856, and 880). The stoichiometric samples exhibit minimal changes to the RHEED pattern during growth, with the only apparent change in the clustering occurring in Sample 740 after transitioning to the step-flow growth regime. Meanwhile, the off-stoichiometric samples have abrupt changes in clustering, which suggests an accumulation of defects over the course of the growth due to excess Ti flux.

## IV. RHEED and XPS data for sample 740 on Nb-doped STO

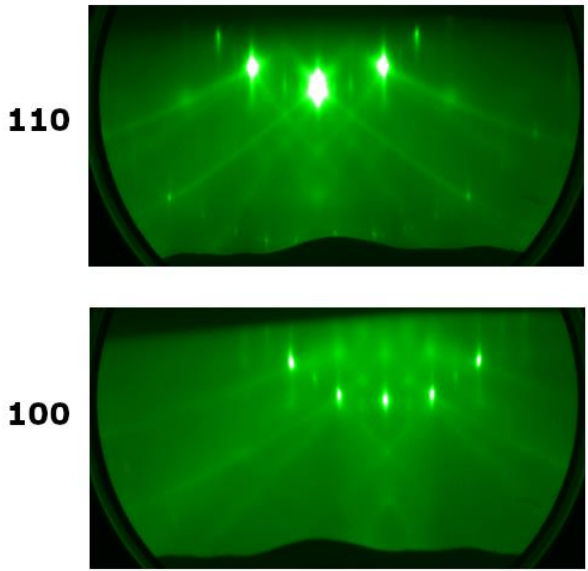

*Figure S3: RHEED patterns along (110) and (100) azimuth for STO film on Nb doped STO substrate.*

Figure S3 is a RHEED pattern for the sample grown on the Nb-doped STO substrate with the same nominal stoichiometry and cation ratio (~0.6) from the XPS analysis as Sample 740. Surface reconstructions are more visible compared to 740 sample along either azimuth. Reconstruction peaks indicate that a c(2×2) reconstruction and (2×2) reconstruction are present, in agreement with the observations by LEED and LEEM.

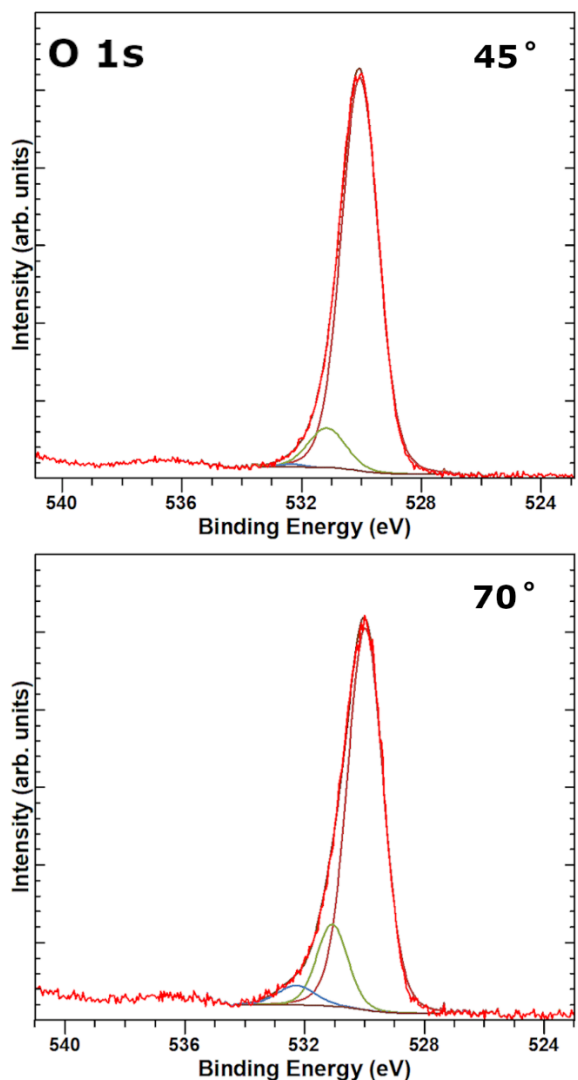

*Figure S4: O1s core level with x-ray incident angle 45° and 70° for STO film on Nb doped STO substrate.*

Figure S4 shows the O 1s the equivalent deconvolution with constraints as described in the main text for the LEEM/LEED sample grown with the same nominal stoichiometry and cation ratio (~0.6) from the XPS analysis as Sample 740 on a Nb-doped STO substrate. As with Samples 740 and 780, this deconvolution shows that the surface oxygen peak at 532.3 eV is more pronounced by relatively surface sensitive orientation of the XPS measurement. The magnitude of the surface component is nearly same as that for the 740 sample. A comparable cation area ratio along with O 1s peak deconvolution reflects same surface chemistry as 740 sample.

## IV. RHEED and XPS data for SrO-terminated STO

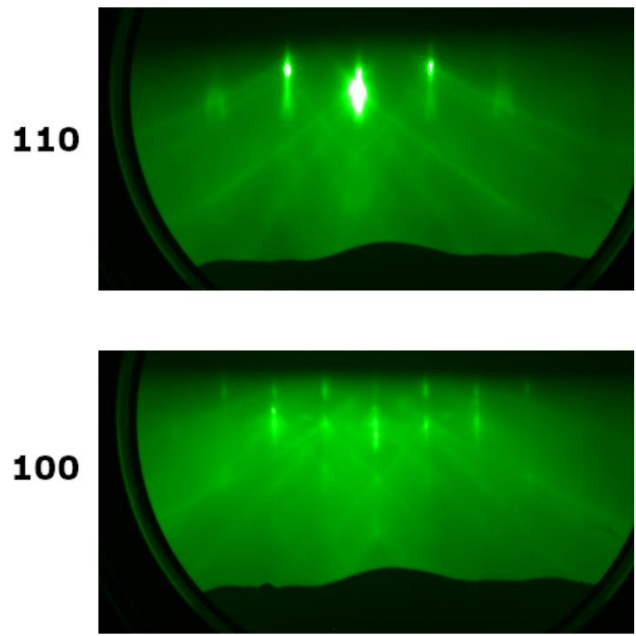

*Figure S5: RHEED patterns along (110) and (100) azimuth for SrO-terminated STO.*

Figure S5 is a RHEED pattern for the SrO terminated STO substrate.

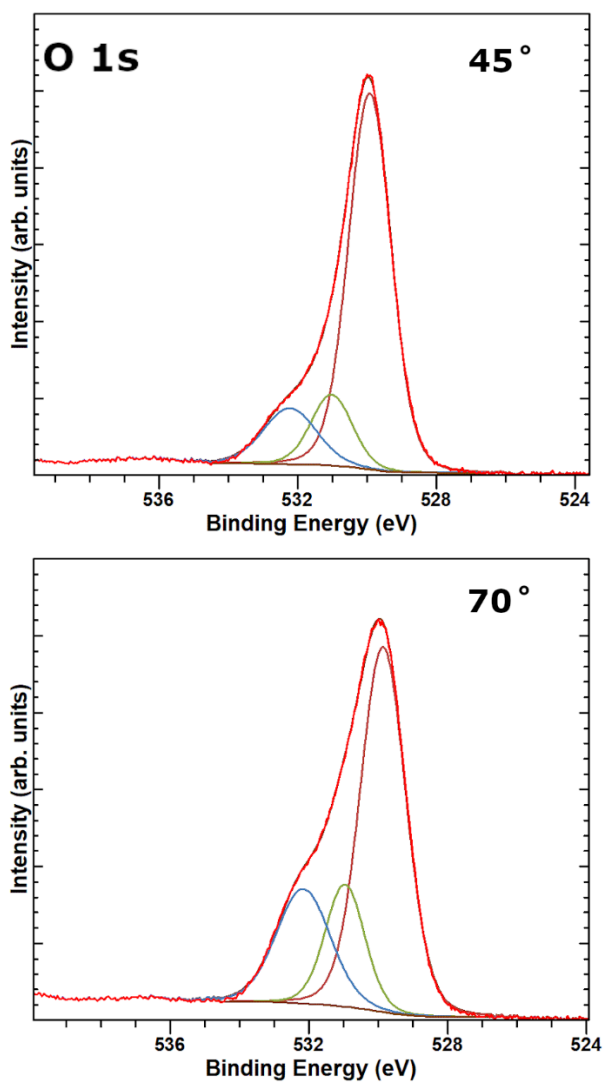

*Figure S6: O1s core level with x-ray incident angle 45° and 70° for SrO-terminated STO.*

Figure S6 shows the O 1s the equivalent deconvolution with constraints as described in the main text for the SrO-terminated STO substrate. As with Samples 740 and 780, this deconvolution shows that the surface oxygen peak at 532.3 eV is more pronounced by relatively surface sensitive orientation of the XPS measurement. A comparable cation area ratio along with O 1s peak deconvolution reflects same surface chemistry as 740 sample.

## V. Angle-resolved XPS Single Crystal Reference Data

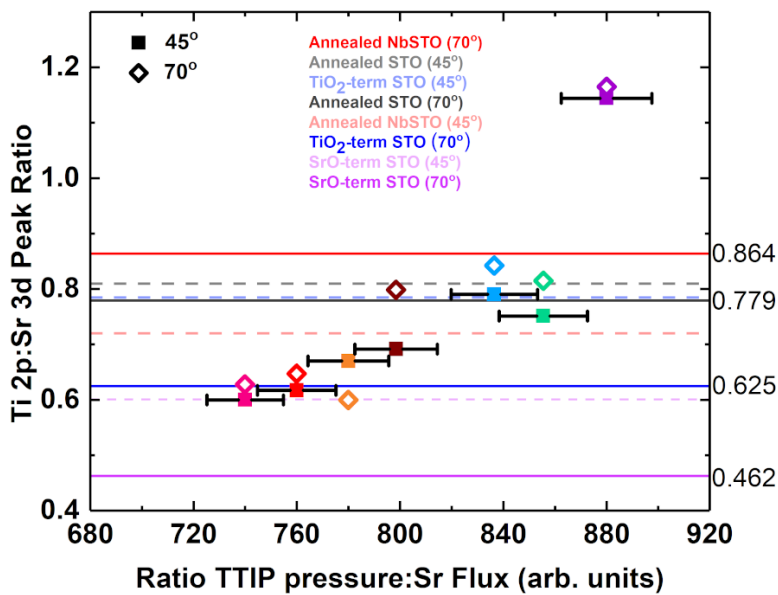

*Figure S7: Replication of Figure 2a including angle-resolved data for single crystal substrates.*

## VI. Ti 2p XPS Comparison

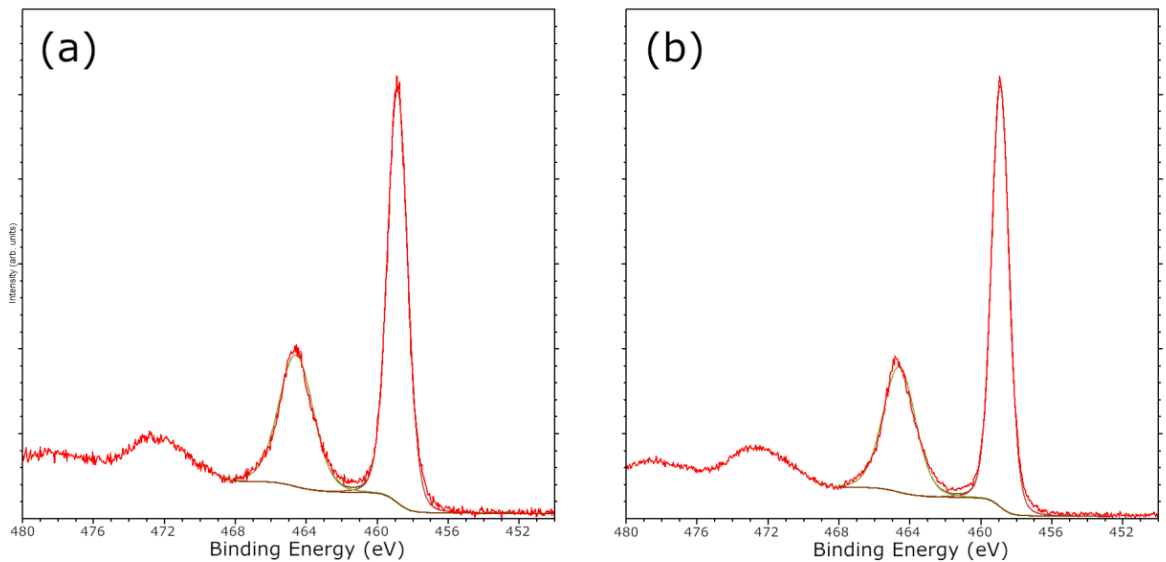

*Figure S8: Comparison of Ti 2p spectra for (a) Sample 798 and (b) Sample 740.*

Figure S8 shows Ti 2p spectra for the off-stoichiometric Sample 798 and stoichiometric Sample 740. Fits to the data by assuming only a $Ti^{4+}$ valence for both samples are essentially identical, suggesting that differences in oxygen vacancy concentrations between samples are not responsible for the differences in XRD patterns. Both spectra were acquired at 45° electron emission angles.

## VII. SrTiO₃ Films Atomic Force Microscopy

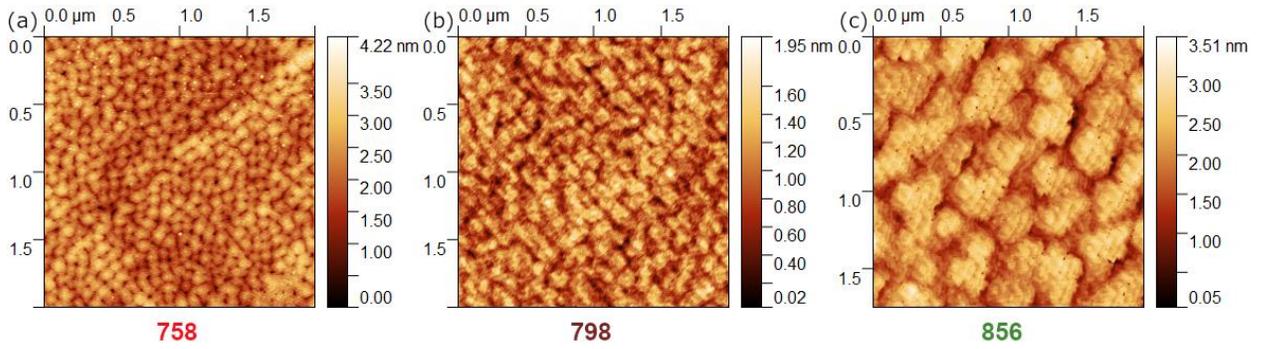

*Figure S9: Atomic force microscopy images of SrTiO₃ homoepitaxial films with flux ratios (a) 758; (b) 798; (c) 856.*

Figure S9 shows atomic force microscopy images of selected SrTiO₃ homoepitaxial films with varying cation flux ratios. The films exhibit similar wedding cake morphologies, albeit with slightly feature size. The morphology is analogous to what is observed by scanning tunneling microscopy reported in the main text.

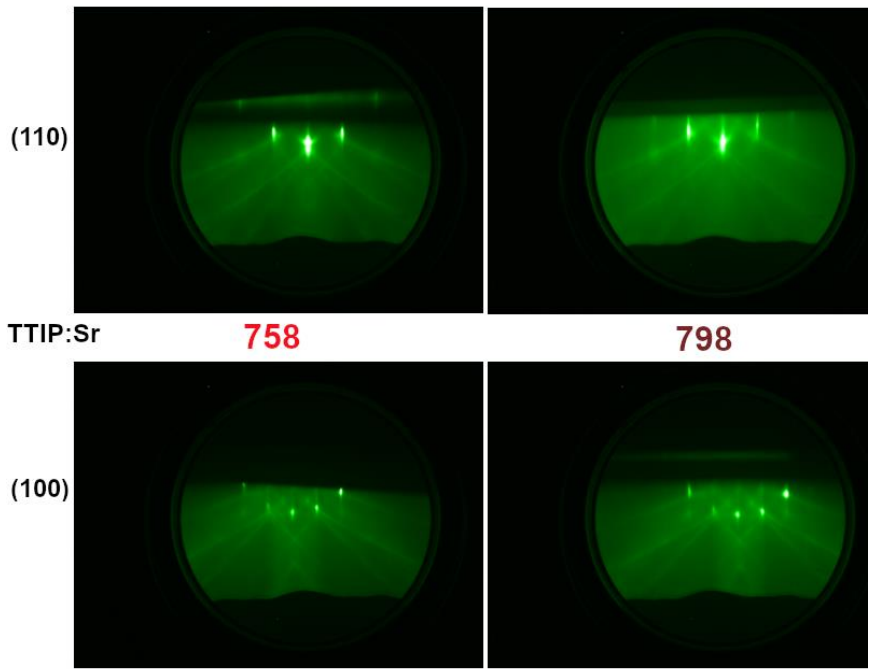

*Figure S10: Atomic force microscopy images of SrTiO₃ homoepitaxial films with flux ratios 758 and 798.*

Figure S10 shows the RHEED patterns for sample 758 and 798.